# An Explainable Physics-Informed Neural Frequency-Response Framework for Shunt-Parameter Identification in Semi-Active Piezoelectric Tuned Mass Dampers

Georgiou A., Gkatsis V., Sioros V., Chatziathanasiou G., Giannakopoulos G.[+], Chrysochoidis N.[+] & Rekatsinas C.[+*]

## Abstract

This paper proposes a Physics-Informed Neural Frequency Response Framework (PI-NFRF) for learning and interpreting the frequency-domain behavior of semi-active shunted piezoelectric tuned mass dampers (*SATMD*). The motivation is that the behavior of such systems is most naturally expressed through frequency response functions, while the governing electromechanical interactions depend strongly on hidden structural and shunt parameters. Conventional data-driven models can approximate these mappings, but they often lack physical consistency, require large training datasets, and provide limited interpretability. To address these limitations, the proposed framework combines a physics-based forward frequency-response model, a neural inverse learning module, and an explainability component.

The forward model is used to generate synthetic complex-valued frequency-response data over a broad range of structural and shunt configurations while preserving the governing electromechanical behavior of the system. Based on synthetic frequency-response data, the neural inverse model is trained to estimate hidden parameters from spectral response signatures and is subsequently evaluated using independently measured experimental FRFs. This synthetic-to-experimental design enables fast parameter inference without solving a new optimization problem for each measured case. To improve robustness to realistic conditions, controlled noise is introduced only at the inverse-training stage, while the underlying physics model remains noise-free. In addition, the learned representation is analyzed through latent-space organization, sensitivity mapping, and reduced symbolic distillation in order to extract interpretable electromechanical response descriptors. The resulting framework provides a data-efficient and explainable machine-learning approach for frequency-response-based identification and inverse tuning of semi-active electromechanical absorbers, while remaining grounded in the validated analytical and experimental SATMD formulation.

*Keywords*

Physics-informed machine learning; Explainable artificial intelligence; Frequency response analysis; Semi-active tuned mass damper; Shunted piezoelectric damping; Inverse identification; Symbolic distillation

*+ Share Last authorship*

** Corresponding Author*

## Introduction

SATMDs offer an attractive route to broadband vibration mitigation in flexible structures by combining a resonant auxiliary mass with electromechanical transduction and reconfigurable shunt impedance.

SATMDs are motivated by a recurring limitation of conventional tuned mass dampers (TMDs): while passive TMDs can be highly effective, they are typically narrowband and sensitive to detuning caused by changes in operating conditions, boundary conditions, or structural parameters. In contrast, an SATMD incorporates a piezoelectric element within the absorber connection, enabling conversion of mechanical vibration energy into electrical energy and shaping the effective absorber dynamics through an external shunt circuit. Foundational studies on piezoelectric shunt damping established that a piezoelectric transducer shunted by passive electrical networks exhibits impedance-dependent, frequency-dependent stiffness and loss behaviour (Hagood and von Flotow, 1991). Resonant RL shunts were subsequently investigated as a means of achieving targeted vibration attenuation by tuning circuit elements to modal dynamics (Wu, 1996), and multimodal strategies were explored to extend attenuation benefits beyond a single resonance (Hollkamp, 1994). Practical realisations have also considered synthetic or adaptive shunt implementations to maintain tuning under changing conditions (Fleming and Moheimani, 2003). These developments collectively highlight that shunted piezoelectric devices are naturally characterised and tuned in the frequency domain, using FRFs as a primary design and validation tool (Gripp and Rade, 2018; Marakakis et al., 2019).

Recent SATMD-specific work has demonstrated that combining a resonant mass with a spring–piezoelectric element and an external resistive–inductive circuit can yield robust, multi-modal vibration suppression with experimentally validated retuning capability (Grigorios M Chatziathanasiou et al., 2022). In particular, SATMD studies report strong sensitivity of structural FRFs to shunt resistance and inductance, including peak attenuation, anti-resonance formation, and modal interaction changes, and show that impedance alterations can recover performance under structural or loading changes (Grigorios M. Chatziathanasiou et al., 2022). While these contributions establish a strong electromechanical modelling and experimental foundation, practical deployment still faces a key challenge: rapid and robust selection (or inference) of shunt parameters from measured frequency-domain signatures, particularly under uncertainties that degrade the reliability of purely analytic tuning rules.

This challenge aligns closely with broader trends in scientific machine learning (SciML), where data-driven models are increasingly combined with physical structure to improve generalisation and interpretability. Physics-informed neural networks (PINNs) embed governing equations into learning objectives and have been widely adopted for forward simulation (Theodosiou and Rekatsinas, 2026) and inverse parameter identification (Raissi et al., 2019) in systems governed by differential equations. Beyond PINNs, theory- and physics-guided approaches blend physics-model outputs with neural networks, or add physics-consistency penalties, to improve scientific plausibility and robustness when data are scarce or imperfect (Karpatne et al., 2017). Neural ODEs introduced continuous-depth modelling by parameterising state derivatives with neural networks and integrating them with ODE solvers, offering a dynamical-systems view of deep learning models (Chen et al., 2018; Lai et al., 2021).

Despite these advances, two limitations constrain the direct application of existing SciML methods to SATMD tuning. First, many physics-informed formulations are time/space-domain centric, whereas SATMD design and assessment is fundamentally frequency-domain: experiments typically report FRFs, tuning decisions are made from spectral signatures, and the governing electromechanical impedance is explicitly frequency dependent. Second, shunt parameters are commonly selected through analytical tuning rules or numerical optimization based on a prescribed electromechanical model (Høgsberg, 2021)[Høgsberg and Krenk, 2012; Zhao et al., 2015; Høgsberg, 2019]. Such procedures generally need to be repeated when the target response or structural state changes [Chatziathanasiou et al., 2022] and may therefore become computationally demanding and sensitive to discrepancies between the nominal model and the actual system.

A related, but equally important, usefulness barrier of recent methods is the lack of explainability. In vibration control, an inferred shunt setting must be trusted not only for numerical accuracy but also for physical plausibility and safe operation. The interpretability literature argues that high-stakes decisions benefit from transparent models rather than opaque predictors with post-hoc explanations (Rudin, 2019). In scientific discovery settings, interpretability is sometimes achieved by extracting parsimonious governing relations from data, for example through sparse regression (SINDy) or symbolic regression methods (Brunton et al., 2016) that search for compact analytical expressions (Kissas et al., 2024; Udrescu and Tegmark, 2020). For SATMDs, these ideas suggest a promising direction: the learning system should not only infer RL parameters from FRFs but also expose which spectral regions and internal descriptors correspond to electromechanical regimes such as peak splitting, anti-resonance movement, or coupling-driven response changes.

Motivated by these gaps, this paper proposes PI-NFRF, a frequency-domain, physics-informed, and explainable learning framework tailored to SATMDs. The framework is built around a differentiable electromechanical frequency-response solver that generates physically consistent synthetic FRFs over a range of shunt configurations and supports direct calibration of physically interpretable model parameters. Its primary ML component is a neural inverse model trained to infer shunt resistance and inductance from complex acceleration FRFs. By reintroducing the predicted parameters into the forward solver during training, the inverse model is constrained by its ability to reconstruct the observed spectral response, enabling amortized inference without solving a separate optimization problem for each new FRF. Synthetic-to-experimental transfer is evaluated using independently measured FRFs from shunt configurations excluded from model development.

The learned behaviour is subsequently examined through three complementary analyses: feature-space organisation, to assess whether internal representations vary systematically across tuning conditions; frequency-wise neural attribution and comparison with physical sensitivities, to identify the spectral regions most influential for parameter inference; and symbolic approximation of a reduced FRF descriptor, to obtain a compact representation of the response variation. Through this hierarchy, PI-NFRF combines physics-based parameter calibration, inverse learning, and interpretable analysis to provide accurate shunt-parameter identification and engineering insight into the spectral mechanisms governing semi-active shunted piezoelectric vibration mitigation.

## Methodology

The framework is formulated directly in the frequency domain because the principal observable is the frequency response function, through which the influence of the resistive–inductive shunt appears as shifts in resonance and anti-resonance locations, changes in damping, and variations in electromechanical coupling. In the SATMD, the piezoelectric element connects the host structure and auxiliary mass, while the external circuit regulates the conversion, storage, and dissipation of vibration energy, enabling tunable multi-modal control. Within this setting, PI-NFRF combines a differentiable electromechanical solver with a neural inverse model that estimates shunt resistance and inductance from complex acceleration FRFs, together with complementary analyses of the learned feature space, frequency-wise attribution, and reduced response descriptors. The framework is physics-informed because the predicted parameters are reintroduced into the governing SATMD model and evaluated according to their ability to reconstruct the observed FRFs, rather than being assessed solely against parameter labels.

### *Frequency-domain model of the SATMD*

The starting point of the framework is the coupled electromechanical model of the SATMD. The host structure is represented by a primary mass, the absorber by a secondary mass, and the piezoelectric connection by an equivalent mechanical stiffness, damping contribution, capacitance, and electromechanical coupling term. The shunt circuit introduces a resistive-inductive branch that gives rise to an additional electrical degree of freedom. In the original SATMD formulation, this leads to a three-degree-of-freedom electromechanical system in which the response is strongly affected by the shunt resistance and inductance. In the frequency domain, the coupled equilibrium can be written as

$$\mathbf{Z}(\omega;\boldsymbol{\theta})\,\mathbf{x}(\omega;\boldsymbol{\theta}) = \mathbf{f}(\omega) \qquad (1)$$

where $\omega$ is the excitation frequency, $\mathbf{x}$ is the complex response vector, $\mathbf{f}$ is the force vector, and $\mathbf{Z}$ is the complex dynamic stiffness matrix. The parameter vector is

$$\boldsymbol{\theta} = [m_2,\, k_p,\, R,\, L] \qquad (2)$$

where $m_2$ is the absorber mass, $k_p$ is the equivalent piezoelectric stiffness, and $R$ and $L$ are the shunt resistance and inductance. In the present implementation, the response vector is defined as

$$\mathbf{x}(\omega;\boldsymbol{\theta}) = \begin{bmatrix} U_1(\omega) \\ U_2(\omega) \\ Q(\omega) \end{bmatrix} \qquad (3)$$

where $U_1$ is the complex displacement of the main structure, $U_2$ is the complex displacement of the auxiliary mass, and $Q$ is the electrical response variable associated with the shunt circuit. Using the notebook implementation, the dynamic stiffness matrix takes the form

$$\mathbf{Z}(\omega;\boldsymbol{\theta}) = \begin{bmatrix} k_1 + k_p - m_1\omega^2 + i\omega(c_1 + c_p) & -k_p - i\omega c_p & e/C_p \\ -k_p - i\omega c_p & k_p - m_2\omega^2 + i\omega c_p & -e/C_p \\ e/C_p & -e/C_p & 1/C_p - L\omega^2 + i\omega R \end{bmatrix} \quad (4)$$

where $m_1$is the primary mass, $k_1$is the structural stiffness, $c_1$is the structural damping, $c_p$is the equivalent damping contribution of the piezoelectric device, $e$is the electromechanical coupling coefficient, and $C_p$is the capacitance of the piezoelectric transducer. The force is applied to the structural degree of freedom, such that

$$\mathbf{f}(\omega) = \begin{bmatrix} F_0 \\ 0 \\ 0 \end{bmatrix} \quad (5)$$

and for each sampled frequency, the complex response is obtained by solving the linear system in (6). The implementation does not form the matrix inverse explicitly; instead, it applies a differentiable complex-valued linear solver.

$$\mathbf{x}(\omega;\boldsymbol{\theta}) = \mathbf{Z}^{-1}(\omega;\boldsymbol{\theta})\,\mathbf{f}(\omega). \quad (6)$$

Equation (6) defines the forward frequency-response model of the coupled electromechanical system. In the updated implementation, the structural acceleration FRFs are also computed explicitly from the displacement response as

$$a_1(\omega;\boldsymbol{\theta}) = -\omega^2 u_1(\omega;\boldsymbol{\theta}),\ a_2(\omega;\boldsymbol{\theta}) = -\omega^2 u_2(\omega;\boldsymbol{\theta}). \quad (7)$$

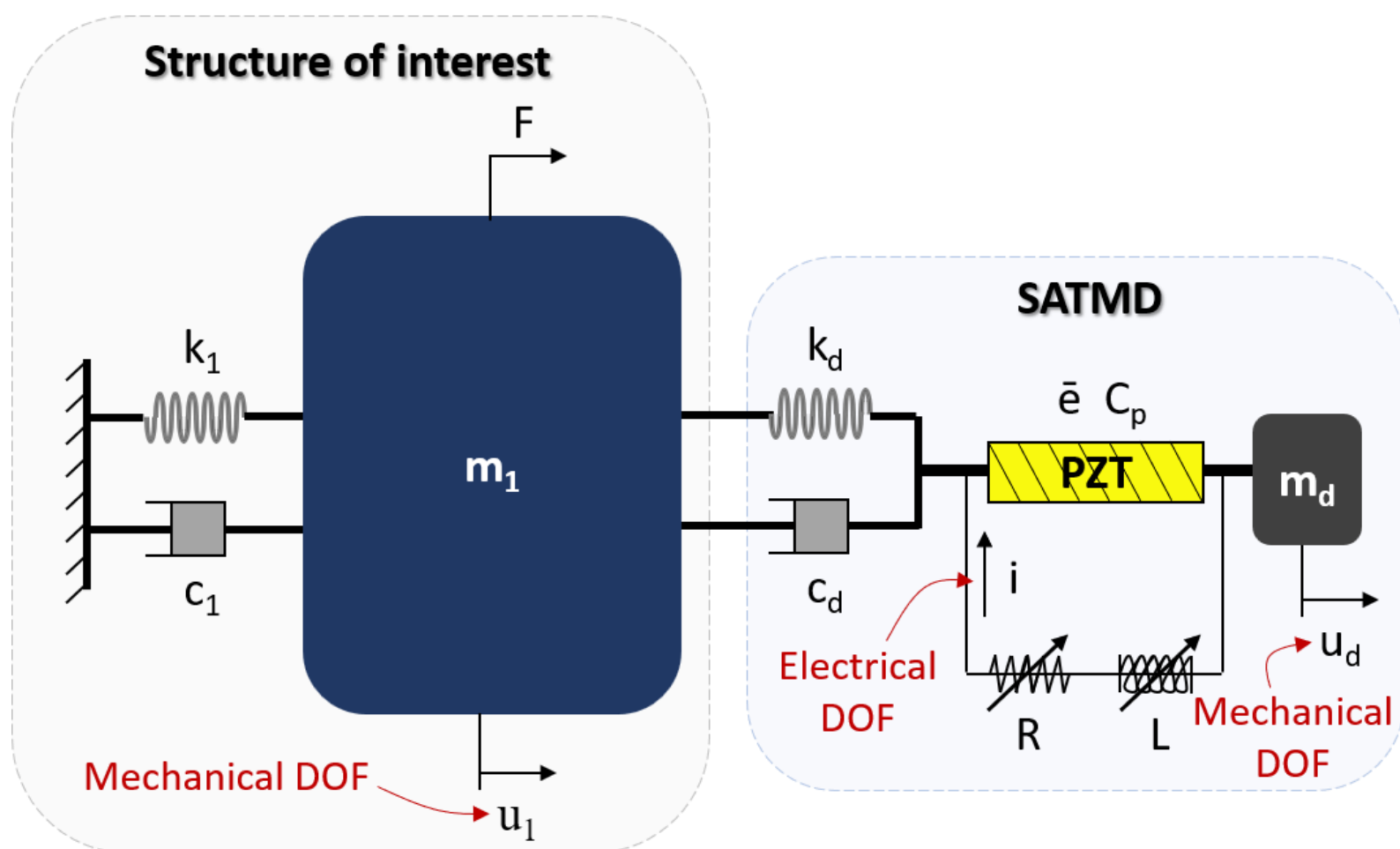

*Figure 1 Schematic of the semi-active shunted piezoelectric tuned mass damper, showing the primary structure, auxiliary mass, piezoelectric transducer, and external (RL) shunt circuit. The acceleration responses ($a_1$) and ($a_2$) are used as inputs to the PI-NFRF framework.*

### *Use Case I: Verification of differentiable frequency-domain parameter identification*

The first use case verifies the differentiable parameter-identification capability of PI-NFRF under controlled conditions. A reference complex FRF is generated using a known electromechanical parameter set, while the identification procedure is initialized from deliberately perturbed nominal values. The absorber mass $m_2$, piezoelectric stiffness $k_p$, shunt resistance $R$, and shunt inductance $L$are then recovered by minimizing the complex-valued FRF reconstruction loss. This experiment evaluates whether differentiation through the frequency-domain solver can recover the reference parameters while preserving their direct physical interpretation.

The first block of PI-NFRF is the forward physics-informed response model, denoted by

$$\mathbf{x}(\omega;\boldsymbol{\theta}) = \mathcal{F}_{\text{phys}}(\omega,\boldsymbol{\theta}). \quad (8)$$

This block is not a purely data-driven model. It is a differentiable physics-based solver that computes the complex frequency response at each sampled frequency. Because it is differentiable, it can also be embedded in optimization and learning loops. Over a discrete frequency grid $\{\omega_j\}_{j=1}^{N_\omega}$ the full spectral signature of the system is written as

$$\mathbf{X}(\boldsymbol{\theta}) = \left[\mathbf{x}(\omega_1;\boldsymbol{\theta}), \mathbf{x}(\omega_2;\boldsymbol{\theta}), \ldots, \mathbf{x}(\omega_{N_\omega};\boldsymbol{\theta})\right]. \qquad (9)$$

This quantity represents the complex frequency-response signature used throughout the framework. For the inverse stage, the implementation uses the structural acceleration signature

$$\mathbf{A}(\boldsymbol{\theta}) = \left[\mathbf{a}(\omega_1;\boldsymbol{\theta}), \mathbf{a}(\omega_2;\boldsymbol{\theta}), \ldots, \mathbf{a}(\omega_{N_\omega};\boldsymbol{\theta})\right], \mathbf{a}(\omega;\boldsymbol{\theta}) = \begin{bmatrix} A_1(\omega;\boldsymbol{\theta}) \\ A_2(\omega;\boldsymbol{\theta}) \end{bmatrix}. \qquad (10)$$

This distinction is important. The forward solver remains capable of returning the full electromechanical state, but the inverse network is trained on the mechanical acceleration response, which is the quantity most naturally linked to measured vibration data.

The first use case addresses parameter identification through direct fitting of the forward electromechanical model to a target frequency-response curve. In the updated implementation, the unknown quantities are written as relative corrections around nominal values, namely

$$m_2 = m_{2,\text{nom}}(1+\delta_m), k_p = k_{p,\text{nom}}(1+\delta_k), \qquad (11)$$
$$R = R_{\text{nom}}(1+\delta_R), L = L_{\text{nom}}(1+\delta_L), \qquad (12)$$

where $\delta_m$, $\delta_k$, $\delta_R$, and $\delta_L$ are trainable correction factors. This parameterization constrains the search around physically meaningful baseline values and improves interpretability. Let $\mathbf{X}^*$ denote the target complex FRF signature and $\widehat{\mathbf{X}} = \mathbf{X}(\boldsymbol{\theta})$ the model prediction. Since the response is complex-valued, the loss is computed on both real and imaginary parts:

$$\mathcal{L}_{\text{id}} = \frac{1}{N_\omega}\sum_{j=1}^{N_\omega}\left[\parallel \frac{\text{Re}(\hat{\mathbf{x}}_j) - \text{Re}(\mathbf{x}_j^*)}{s} \parallel_2^2 + \parallel \frac{\text{Im}(\hat{\mathbf{x}}_j) - \text{Im}(\mathbf{x}_j^*)}{s} \parallel_2^2\right] + \lambda_{\text{reg}} \parallel \boldsymbol{\delta} \parallel_2^2 \qquad (13)$$

where $s$ is a normalization factor taken as the maximum magnitude of the target response, and

$$\boldsymbol{\delta} = [\delta_m, \delta_k, \delta_R, \delta_L]. \qquad (14)$$

This first use case remains fully physics-driven. The trainable variables are not arbitrary latent coefficients, but physically interpretable corrections to the electromechanical model parameters.

### *Use Case II: inverse learning from structural acceleration FRFs*

The second use case is the main inverse learning task. In the updated implementation, the inverse network is trained to infer the shunt parameters $R$ and $L$from the **complex structural acceleration response** of the system.

A synthetic dataset is generated by sampling multiple shunt settings

$$\boldsymbol{\theta}^{(n)} = [R^{(n)}, L^{(n)}], n = 1, \ldots, N_s \qquad (15)$$

while keeping the remaining mechanical parameters fixed. For each pair, the forward physics model computes the acceleration signature

$$\mathbf{A}^{(n)} = \mathcal{F}_{\text{acc}}(\omega; R^{(n)}, L^{(n)}), \qquad (16)$$

where $\mathcal{F}_{\text{acc}}$ denotes the acceleration-producing branch of the forward solver. In the code, each sampled pair is further perturbed within a local neighborhood of predefined $(R, L)$ reference parameter pairs, which creates a denser training cloud and improves the local smoothness of the inverse map. To prevent information leakage, the dataset is partitioned at the level of the reference configurations rather than at the level of individual perturbed samples. All samples generated around a given $(R, L)$ pair are assigned to the same training, validation, or test partition. Consequently, no local neighborhood associated with a validation or test reference pair is represented in the training set.

The inverse model does not directly process complex tensors. Instead, the complex acceleration response is converted into a real-valued feature vector by concatenating the real and imaginary parts over the full frequency grid:

$$\mathbf{g}^{(n)} = [\text{Re}(\mathbf{A}^{(n)}), \text{Im}(\mathbf{A}^{(n)})]. \qquad (17)$$

Thus, the neural network input is a compact spectral encoding of the two acceleration FRFs.
The inverse model learns the mapping

$$[\hat{R}, \hat{L}] = \mathcal{M}_{\text{inv}}(\mathbf{g}), \qquad (18)$$

but, in the current implementation, it first predicts normalized variables in the interval $[0, 1]$, which stabilizes training and keeps predictions within the admissible parameter range.

$$[\hat{r}, \hat{\ell}] = \mathcal{N}_{\text{inv}}(\mathbf{g}), \hat{r}, \hat{\ell} \in [0,1], \qquad (19)$$

which are then denormalized to physical values through the corresponding parameter bounds, thereby recovering resistance and inductance in their original units for physics-based FRF reconstruction and direct comparison with the reference circuit settings.

$$\hat{R} = R_{\min} + \hat{r}(R_{\max} - R_{\min}), \hat{L} = L_{\min} + \hat{\ell}(L_{\max} - L_{\min}). \qquad (20)$$

At present, $\mathcal{M}_{inv}$ has been implemented using a fully connected feed-forward neural network. The input for the network is the vector $g$ which contains the spectral features vectorized and has size 240, since there are two accelerometers, 60 ampled frequency points, and real and imaginary parts. The fully connected neural network has three hidden layers whose widths are 256, 128, and 64, respectively, followed by an output layer whose dimension is two. The output layer uses the sigmoid activation function because the shunt parameters are constrained to the interval $[0,1]$.

The inverse model is **not trained only by direct regression on parameter labels**. Instead, the predicted parameters are reintroduced into the forward electromechanical model, and the resulting acceleration FRF is compared with the target acceleration FRF. For a training sample $n$, the reconstructed response is

$$\hat{\mathbf{A}}^{(n)} = \mathcal{F}_{\text{acc}}(\omega; \hat{R}^{(n)}, \hat{L}^{(n)}). \qquad (21)$$

The inverse loss is then defined as a spectral reconstruction loss on the acceleration response:

$$\mathcal{L}_{\text{inv}} = \frac{1}{N_s} \sum_{n=1}^{N_s} \frac{1}{N_\omega} \sum_{j=1}^{N_\omega} \left[ \| \frac{\text{Re}(\hat{\mathbf{a}}_j^{(n)}) - \text{Re}(\mathbf{a}_j^{(n)})}{s_n} \|_2^2 + \| \frac{\text{Im}(\hat{\mathbf{a}}_j^{(n)}) - \text{Im}(\mathbf{a}_j^{(n)})}{s_n} \|_2^2 \right] \qquad (22)$$

where $s_n$ is a sample-wise normalization factor taken as the maximum amplitude of the target acceleration response for the corresponding sample. This is an important difference from the earlier formulation. The inverse neural network is now trained in a **physics-consistent closed loop**, where the predicted $R$ and $L$ are accepted only if they reconstruct the observed spectral behavior through the governing electromechanical model.

The updated framework also preserves the distinction between clean physics and noisy learning data. Whne noise is introduced, it is added to the inverse input signatures,

$$\tilde{\mathbf{g}}^{(n)} = \mathbf{g}^{(n)} + \boldsymbol{\eta}^{(n)}, \qquad (23)$$

while the forward electromechanical solver itself remains noise-free. This ensures that measurement uncertainty is learned at the inference level without contaminating the governing physical model. After training, the inverse network is evaluated using independently acquired experimental acceleration FRFs. The predicted shunt parameters are compared with reference resistance and inductance values obtained from the corresponding circuit configurations.

$$\varepsilon_R = | \hat{R} - R |, \varepsilon_L = | \hat{L} - L |. \qquad (24)$$

This evaluation is performed on blind parameter combinations (randomly selected experimental measurments) that are not used during training, in order to evaluate interpolation across the held-out configurations within the sampled parameter region.

*Use Case III: explainable discovery of electromechanical response descriptors*

The third use case focuses on explainability. The aim is not to claim recovery of the full governing equations from data, but to understand how the trained model organizes electromechanical behavior and to extract reduced interpretable descriptors. If the inverse network includes an encoding stage, each response signature can be mapped to a latent representation

$$\mathbf{z} = \mathcal{E}(\mathbf{X}), \qquad (25)$$

where $\mathcal{E}$ denotes the encoder. The first explainability objective is to determine whether the latent space is organized according to physically meaningful electromechanical regimes. In the present problem, such regimes may correspond to similar anti-resonance locations, similar resistance-inductance tuning states, similar modal peak splitting, or similar levels of electrical-mechanical coupling. Since the SATMD exhibits pronounced spectral changes under variation of $R$ and $L$, a structured latent organization would indicate that the learned representation varies continuously and systematically across the sampled FRF tuning configurations. This is important because it suggests that the network captures physically meaningful electromechanical trends, rather than merely memorizing individual response curves.

The second explainability objective is to identify which frequency regions are most influential in the inverse predictions. This can be quantified through local sensitivities of the predicted parameters with respect to the input spectral signature:

$$S_R(\omega_j) = \left| \frac{\partial \hat{R}}{\partial \mathbf{X}(\omega_j)} \right|, S_L(\omega_j) = \left| \frac{\partial \hat{L}}{\partial \mathbf{X}(\omega_j)} \right| \qquad (26)$$

These sensitivities indicate which parts of the spectrum carry the most information for parameter inference. This is physically important because not all frequency regions are equally informative. For example, frequency bands associated with peak splitting, anti-resonance shifts, or strong electromechanical interaction are expected to play a dominant role in identifying the shunt settings.

The final explainability step is reduced symbolic distillation. Rather than attempting to reproduce the complete high-dimensional, complex-valued FRF with a single analytical expression, the spectral response is first summarized through a physically meaningful scalar descriptor that quantifies the relative deviation of a parameterized FRF from a nominal reference response. Symbolic regression is then applied to express this reduced descriptor explicitly in terms of the underlying physical parameters. The descriptor therefore provides the reduced representation of the FRF variation, while symbolic distillation yields the compact analytical relation used to interpret that variation, without claiming to recover the full governing equations or replace the forward electromechanical solver. The general form of the descriptor is

$$\xi_{FRF} = \frac{\left\| Y(R, L, m_2, k_p) - Y_{nom} \right\|_2}{\left\| Y_{nom} \right\|_2}, \quad (27)$$

Here, $Y(R,L,m_2,k_p)$ is used to denote the complex-valued frequency response vector from the forward SATMD model using some specified physical parameter values, whereas $Y_{nom}$ represents the fixed nominal frequency response. The numerator represents the overall difference between the frequency responses of the sample and nominal models on the frequency mesh, while the denominator scales the difference with respect to the norm of the nominal frequency response. Hence, the measure $\xi_{FRF}$ quantifies this deviation in a relative scalar fashion. Therefore, the problem of symbolic distillation is expressed as the following fitting:

$$\xi_{FRF} \approx f_{SR}(R, L, m_2, k_p), \quad (28)$$

where $f_{SR}$ is a symbolic expression resulting from symbolic regression analysis. The proposed distillation gives an engineering level representation of the FRF variation in terms of the physical parameters considered. It neither substitutes the forward solver nor attempts to derive the underlying exact equations governing the dynamics of the problem. Instead, it provides a concise scalar description of a complicated spectral response.

The complete workflow of PI-NFRF (Figure 2) can therefore be summarized as follows. First, the SATMD is modeled in the frequency domain through a coupled dynamic stiffness formulation. Second, this model is implemented as a differentiable forward response solver and parameterized through physically meaningful corrections around nominal values. Third, the solver is used for direct parameter identification and for generation of synthetic FRFs over a range of shunt and structural parameters. Fourth, an inverse learning model is trained on these synthetic responses, optionally corrupted with controlled noise, to infer hidden shunt settings from spectral measurements. Fifth, the trained inverse model is tested on blind experimental FRFs. Finally, the learned internal representation is interrogated through latent analysis, sensitivity mapping, and reduced symbolic distillation in order to reveal interpretable electromechanical descriptors.

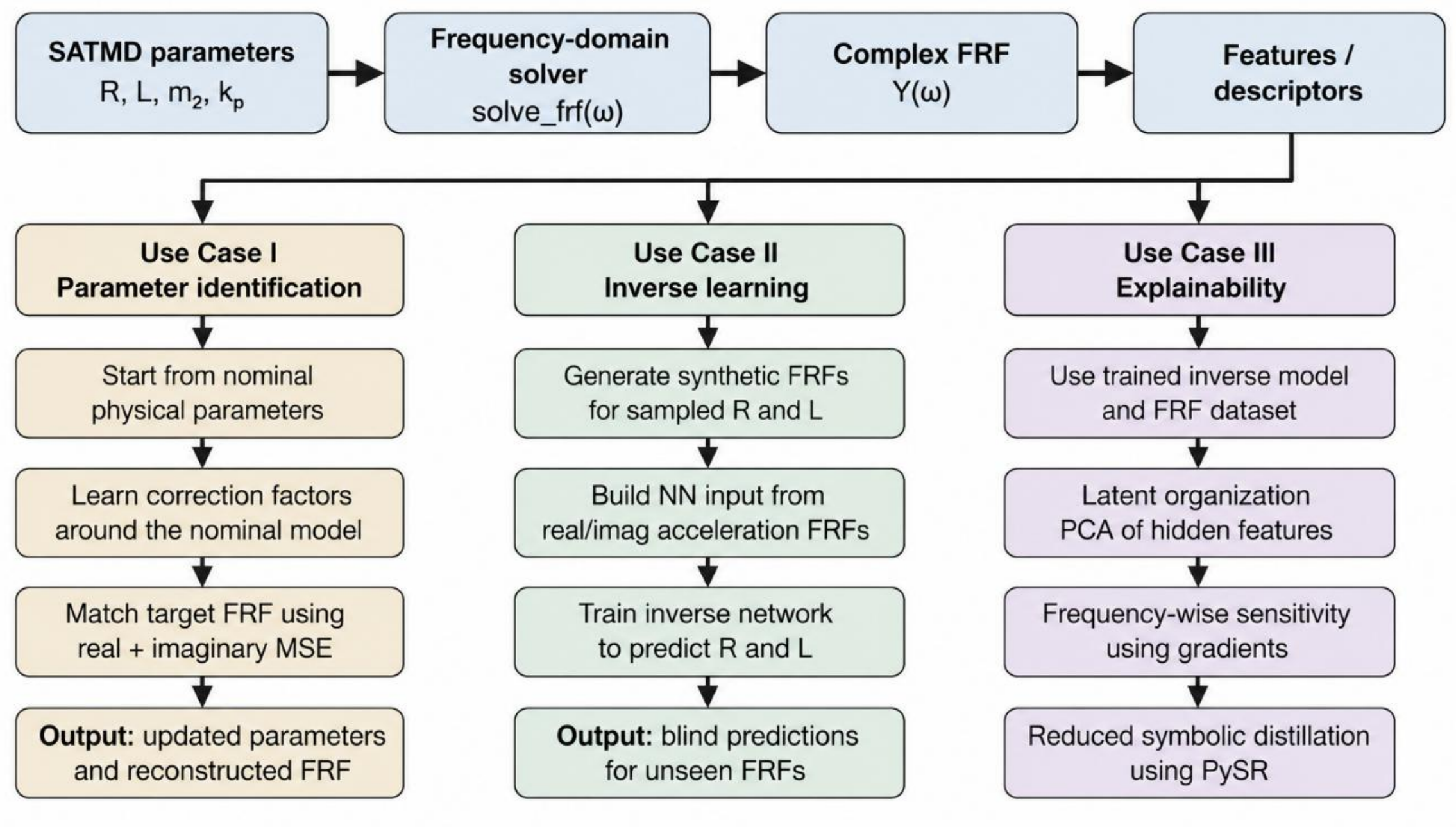

*Figure 2 Neural FRF solver – identification, inverse learning, and explainable analysis*

In this way, the proposed methodology combines electromechanical physics, inverse learning, and explainability within a single frequency-domain framework for semi-active shunted piezoelectric tuned mass dampers and related coupled electromechanical systems.

# Results and discussion

## Use Case I — Physics-informed parameter identification from experimental FRFs

### *Identification setup and nominal model*

The first experiment evaluates the fully physics-informed identification mode of the PI-NFRF framework. The target FRF was fitted by optimizing four physically interpretable parameters of the coupled electromechanical model: the absorber mass $m_2$, the equivalent piezoelectric stiffness $k_p$, the shunt resistance R, and the shunt inductance L. The optimization was initialized from the nominal parameter set $m_2$= 0.300, $k_p$= 50,000, R = 20.0, L = 0.050.

Instead of learning an unconstrained black-box mapping, the model estimates corrections around these nominal values, since the goal is keeping the inverse problem physically interpretable. This means that each optimized value can be interpreted as an updated mechanical or electrical parameter of the SATMD model.

### *Identified parameters*

The optimization found the parameter values shown in Table 1. The electrical parameters are still very close to the nominal circuit values, but the mechanical absorber parameters need bigger changes. This means that the nominal electrical model is already a good place to start, but the effective absorber mass and piezoelectric stiffness need to be changed more to match the measured spectral response.

*Table 1 Nominal and identified parameters for Use Case I*

| Parameter | Nominal value | Identified value | Reference Value | Relative change |
|---|---|---|---|---|
| Absorber Mass $m_2$ | 0.300 | 0.517999 | 0.518 | +72.7% |
| Piezoelectric stiffness $k_p$ | 50,000 | 110,999.797 | 111,000 | +122.0% |
| Resistance R | 20.000 | 21.999981 | 22.000 | +10.0% |
| Inductance L | 0.050 | 0.064000 | 0.064 | +28.0% |

### *FRF reconstruction accuracy*

The relative FRF reconstruction error obtained after identification was $2.35 \times 10^{-6}$. This very small error shows that the updated physics-based frequency-response model reproduces the target complex FRF almost exactly. Therefore, the first use case validates the direct parameter-identification stage of the framework: the experimental frequency response can be matched accurately while retaining a transparent set of updated physical parameters. The corresponding magnitude comparison between the target and reconstructed FRFs is shown in Figure 3. The three FRF curves correspond to the three coupled responses in the SATMD model. In the experimental configuration, the system dynamics are not solely defined by the dynamics of the main structure, but also by the dynamics of the absorber mass and the electrical dynamics brought by the presence of the shunted piezoelectric branch. Thus, the frequency-response vector consists of several entries rather than only a single output response.

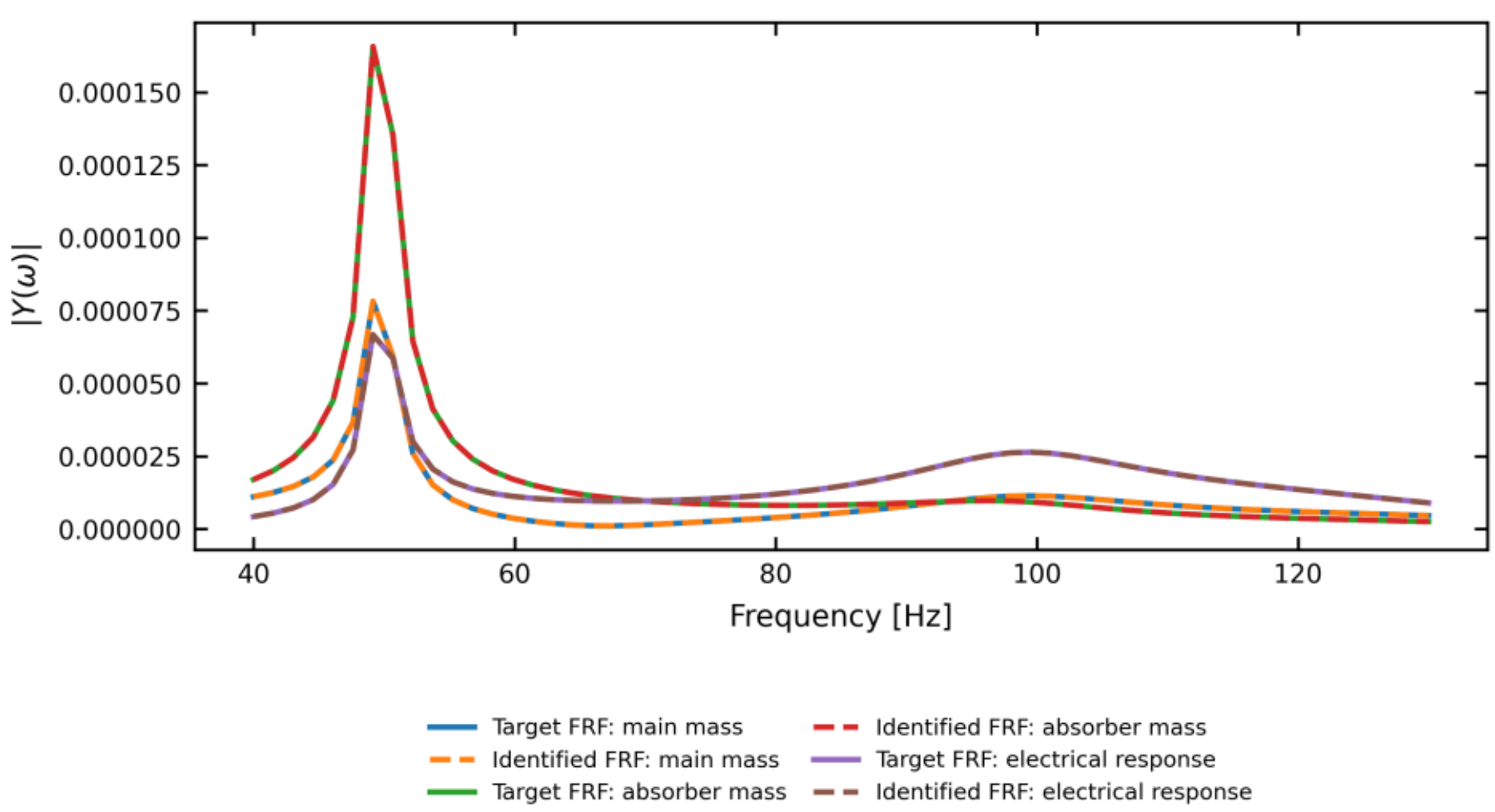

*Figure 3 Frequency-response comparison between the target FRFs and the reconstructed FRFs after physics-informed parameter identification*

## Use Case II — Blind inverse tuning of shunt parameters from experimental frequency response data

In the second scenario, we examine the performance of the proposed PI-NFRF method in terms of inverse learning. In contrast to previous work, here the idea is to estimate shunt parameters (R and L), rather than repeatedly solving the optimization problem for each target FRF.

Inverse training involved acceleration FRFs obtained through physics-based electromechanical forward solving. Every complex response function was represented by a real-valued feature vector, constructed as a concatenation of real and imaginary parts along the entire frequency range. Shunt parameters (R and L) were inferred in normalized form, after which de-normalization yielded physical parameter values. During training, the predicted parameters were reintroduced into the forward model and the loss was computed from the mismatch between reconstructed and target acceleration FRFs. Therefore, the inverse model was trained in a physics-consistent closed loop, where a prediction is useful only if it reproduces the observed spectral behavior through the governing electromechanical model.

To assess generalization, the trained inverse model was evaluated on six blind shunt configurations that were not used as anchor training pairs. The results are summarized in Table 2.

*Table 2 Blind inverse prediction of shunt resistance and inductance from acceleration FRFs*

| $R_{true}$ [Ω] | $L_{true}$ [H] | $R_{pred}$ [Ω] | $L_{pred}$ [H] | $\|e_R\|$ [Ω] | $\|e_L\|$ [H] |
|---|---|---|---|---|---|
| 8.0 | 0.032 | 8.0709 | 0.03154 | 0.0709 | 0.000463 |
| 10.0 | 0.050 | 10.0535 | 0.04956 | 0.0535 | 0.000438 |
| 22.0 | 0.064 | 21.9515 | 0.06358 | 0.0485 | 0.000423 |
| 25.0 | 0.100 | 25.0939 | 0.09981 | 0.0939 | 0.000187 |
| 35.0 | 0.192 | 35.0788 | 0.19126 | 0.0788 | 0.000736 |
| 60.0 | 0.300 | 60.1483 | 0.29881 | 0.1483 | 0.001187 |

For the six blind cases, the neural network yielded an average absolute error of 0.0823 Ω for R and 0.000572 H for L. The mean relative errors are therefore 0.41% for R and 0.66% for L. The highest absolute errors were found for R equal to 60.0 Ω, where the error is 0.1483 Ω, and for L equal to 0.300 H, where the error is 0.001187 H.

This demonstrates that the errors stay small for the entire range tested, including the low R and low L as well as the high R and high L cases. This shows that the acceleration FRF indeed carries enough information for determining the shunt parameters and that the inverse neural network has learned a continuous relationship between the FRFs and the tuning parameters.

## Use Case III — Interpretation of learned electromechanical response representations

### *Latent Organization*

The first explainability approach examines the internal workings of the inverse model to determine the structure of the input acceleration FRFs prior to the estimation of the final shunt parameters. In the current design, the inverse model first maps the FRF inputs into a latent representation form. The latent vector is obtained from the hidden layer that precedes the final output layer of the inverse model. In the current design, the latent vector has a dimension of 64. Therefore, for each FRF sample, the trained model produces a feature vector of the form $z_i \in R^{64}$.

This 64-dimension vector can thus be viewed as the concise representation extracted from the inverse neural network. This vector consists of information deemed critical in predicting shunt resistance and inductance by the model. The output layer is responsible for mapping the latent representation into the normal prediction of the values R and L.

Given that the latent vector is 64 dimensions, it is impossible to visualize directly. It is for this reason that PCA is performed only as an after-processing and visualization technique. Specifically, the latent vectors obtained from all the FRF samples are extracted and mapped onto the first two principal components, $z_i \in R^{64} \rightarrow z_i^{PCA} \in R^2$
As such, the axes indicated on the latent plot charts, labeled as Latent PC1 and Latent PC2, are not the actual neural network outputs but the first two principal components of the 64-dimensional latent feature vectors, which capture the primary directions of variability in the latent space.

A point on the plot represents an FRF for a single acceleration measurement. The location of each point is defined through the PCA transformation of the 64-dimensional latent space vector associated with the respective sample. At the same time, the coloring shows the physical value of the shunt parameter corresponding to this sample.

On the first latent plot, the same points are shown with the coloring based on the resistance R. It can be seen that the projection creates a curved manifold rather than a disordered cluster of points. This means that the inverse problem solver learns a structured representation of the input FRF data. Samples with similar spectral characteristics are grouped close together, whereas points representing various shunt settings are mapped to different parts of the manifold. On the plot with coloring by R, the color changes continuously along the curve, illustrating the informative latent representation of the resistance value.

Similarly, the PCA projection is displayed based on the inductance value L. Note that the point locations in this figure are the same as in the previous figure since both the latent vector components and PCA projection remain the same in each case. The difference lies in the color-coding variable used to plot the graph. This is significant since the two figures above are not illustrating two separate latent spaces, but rather the same latent space projected from two different physical quantities. Again, there is a smooth variation in color values for L, whereby low inductance values are located on one side of the latent manifold, whereas high inductance values progress along the same curve. This implies that the learned 64-dimensional latent vector can capture the inductance information as well.

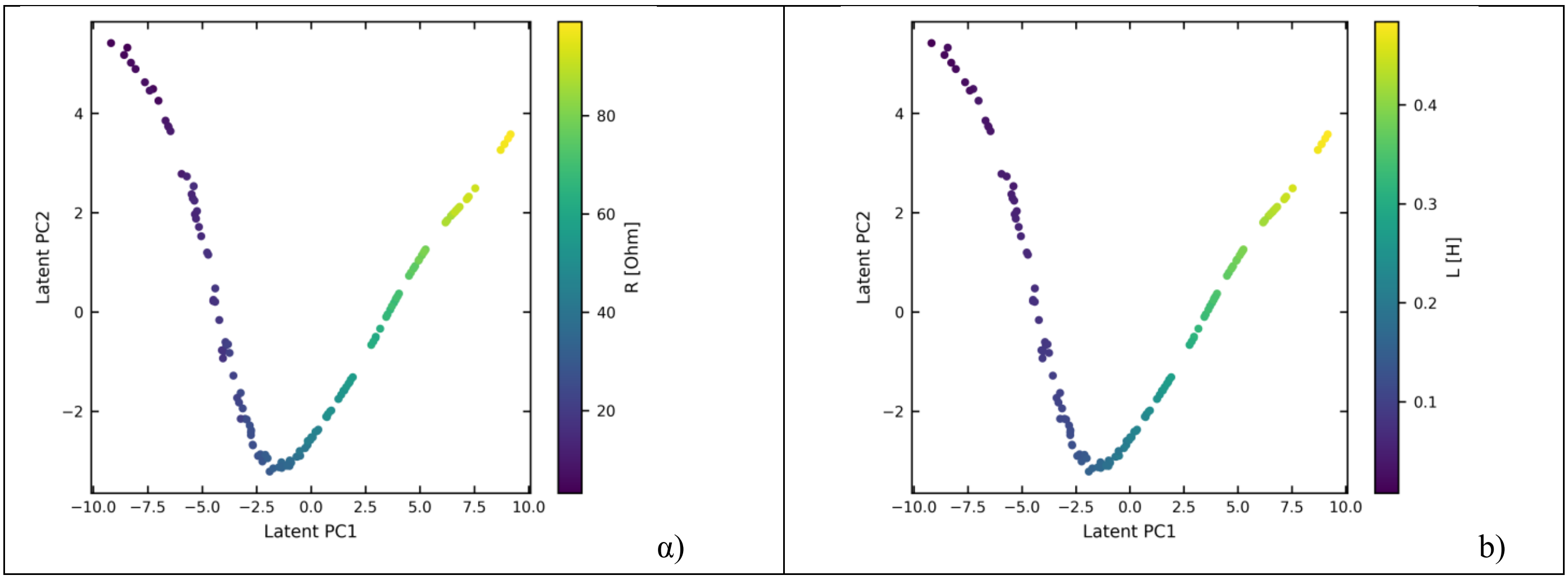

*Figure 4 PCA projection of the 64-dimensional latent features, colored by α) resistance R΄b) inductance L*

However, the latent plots should not be interpreted as proof that the network has fully disentangled R and L into two independent directions. In the blind experimental cases considered here, R and L vary together across the tested configurations. As a result, both parameters change smoothly along the same latent manifold. This suggests that the model has mainly learned a combined electromechanical tuning coordinate, rather than two completely separate latent axes for resistance and inductance. To demonstrate full disentanglement, additional cases would be required where R is varied while L is fixed, and where L is varied while R is fixed.

## Frequency-wise sensitivity to the shunt resistance R

The normalized sensitivity of the estimated resistance with respect to the input acceleration FRF across the entire frequency band explored for each of the six blind shunts is depicted in Figure 5. It is worth noting that the sensitivity does not vary consistently across all frequencies; instead, there are peaks of sensitivity located at certain frequency intervals. Notably, the highest and most pronounced peak occurs at a frequency range of about 50 to 52 Hz, whereas the secondary peak exists within the frequency range of 103 to 110 Hz. It is important to highlight that these specific frequency ranges correspond to regions of the FRF where significant variations can occur in resonance amplification, anti-resonance behavior, and mode interaction.

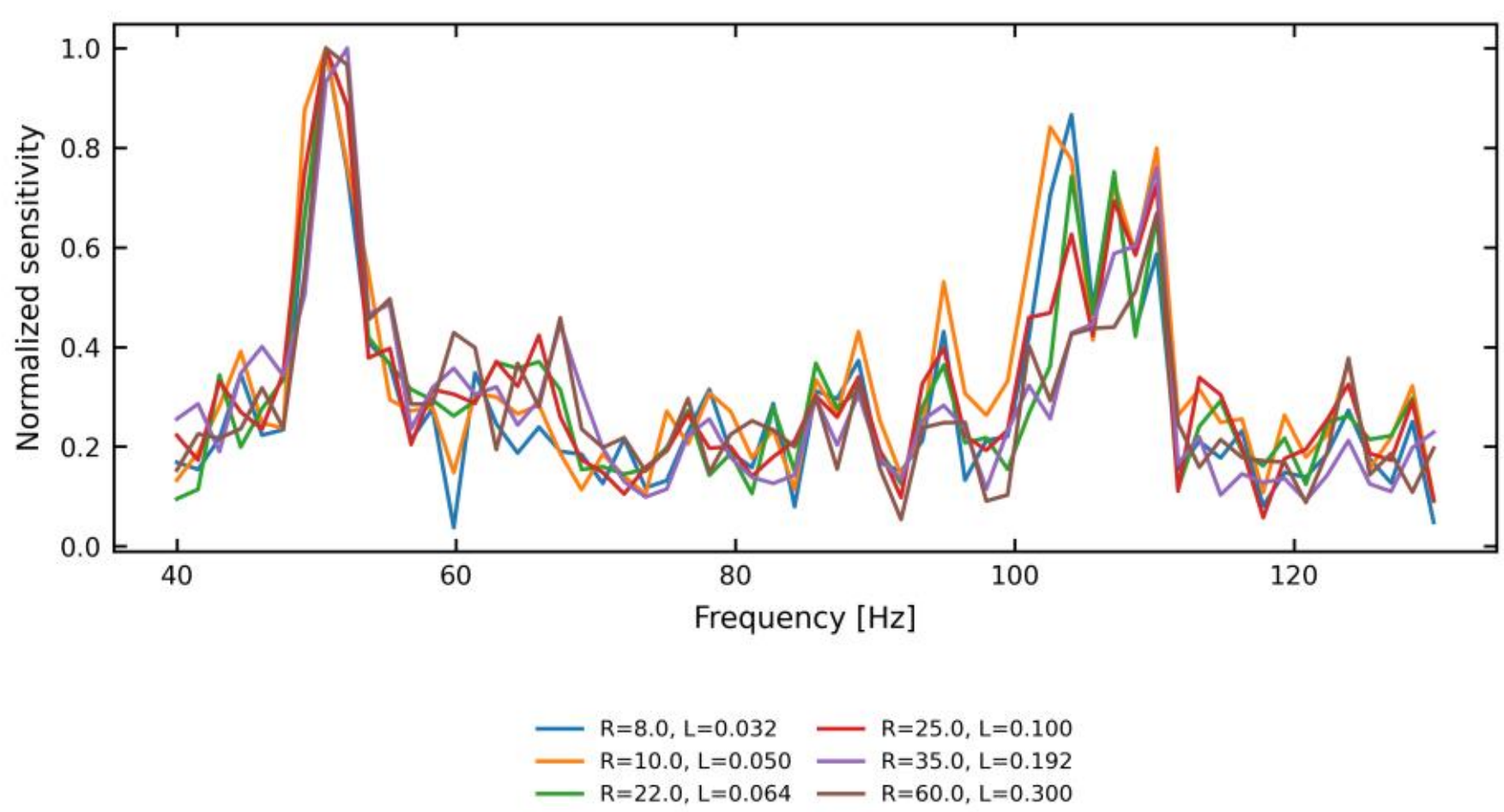

*Figure 5 Normalized sensitivity to R across the six blind shunt configurations*

This same behavior is illustrated in the heatmap presented in Figure 6. The y-axis corresponds to the blind instances, while the x-axis corresponds to the frequency. The bright vertical line on the first resonance zone implies that there exists a common frequency range of information in resistance recognition for all instances under blind conditions. A wider high sensitivity range can be observed in the high frequency zone, at 100 to 110 Hz, although its exact brightness level may vary from instance to instance.

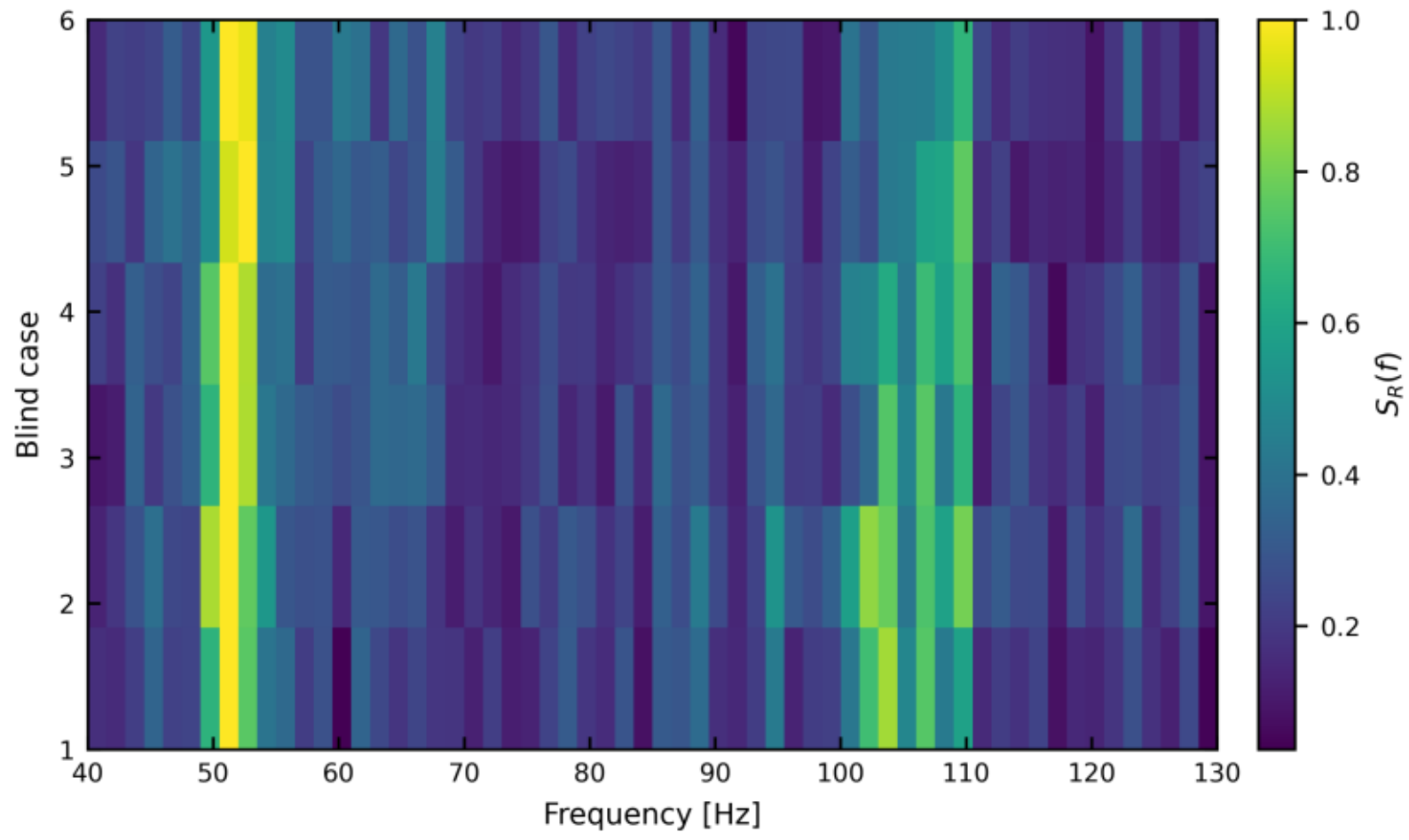

*Figure 6 Heatmap of normalized sensitivity to R across the blind cases*

## Frequency-wise sensitivity to the shunt inductance L

Figure 7 shows the sensitivity curves with respect to inductance. As in the case of resistance, the model gives more importance to a few regions in the frequency domain than to the entire FRF. First of all, the dominant sensitivity region in the case of inductance lies in the vicinity of 50-52 Hz. Second, there is another sensitivity region lying between 103-110 Hz where the normalized sensitivity values of a number of blind cases become high. It should be noted that the sensitivities due to changes in the inductance are higher in this second frequency band due to the function played by inductance in determining the nature of electrical resonance.

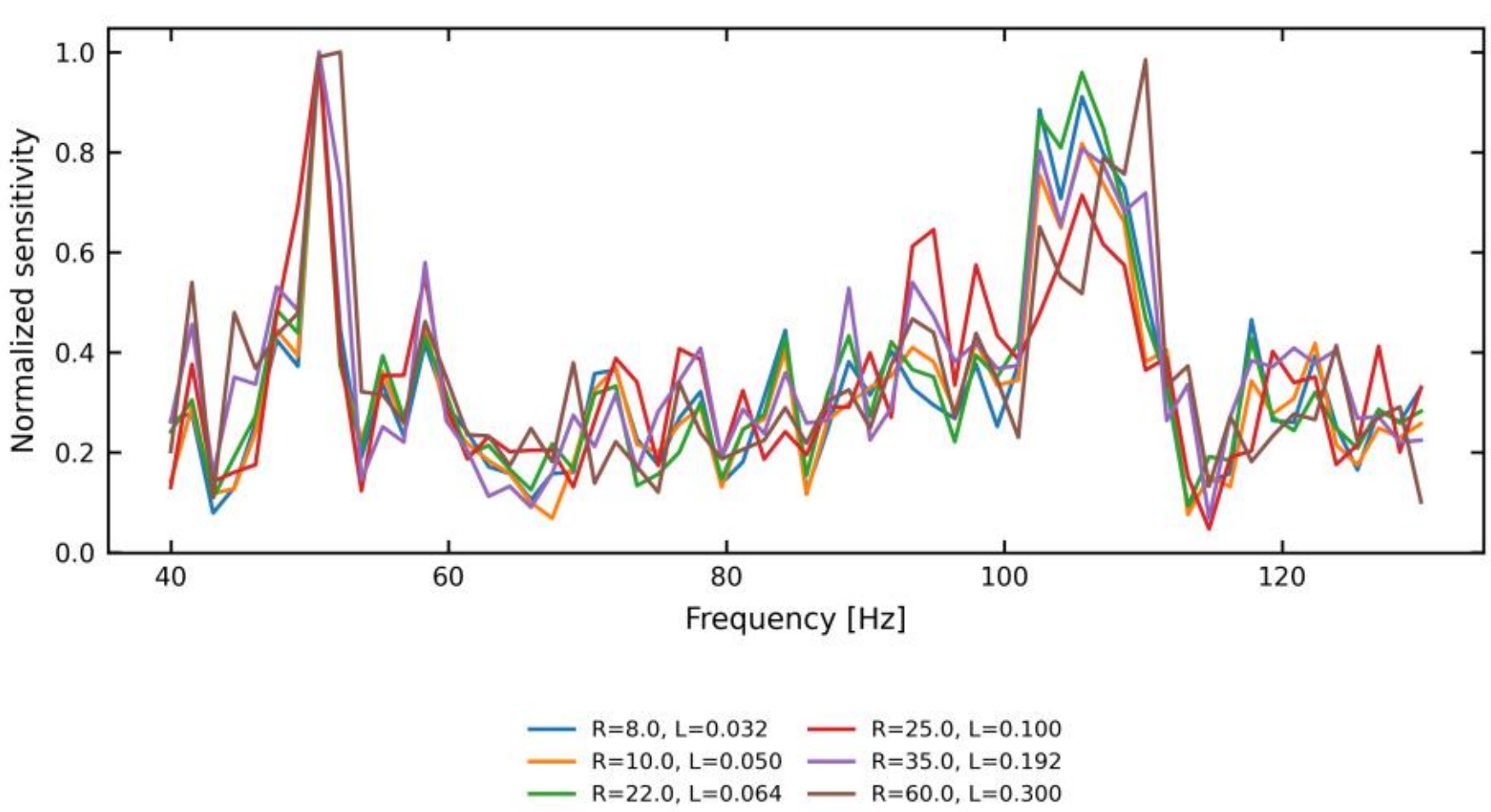

*Figure 7 Normalized sensitivity to L across the six blind shunt configurations*

The heatmap depicted in Figure 8 indeed proves that the most meaningful areas for L coincide in the blind cases to a large extent. In particular, one can observe a highly sensitive area in the range of 50-52 Hz for all cases, whereas a relatively more meaningful area appears in the higher frequencies (103-110 Hz), although the latter exhibits a stronger dependency on the particular case. It should not come as a surprise because the inductance value directly influences the electrical resonance, and thus alters the most meaningful spectral features accordingly.

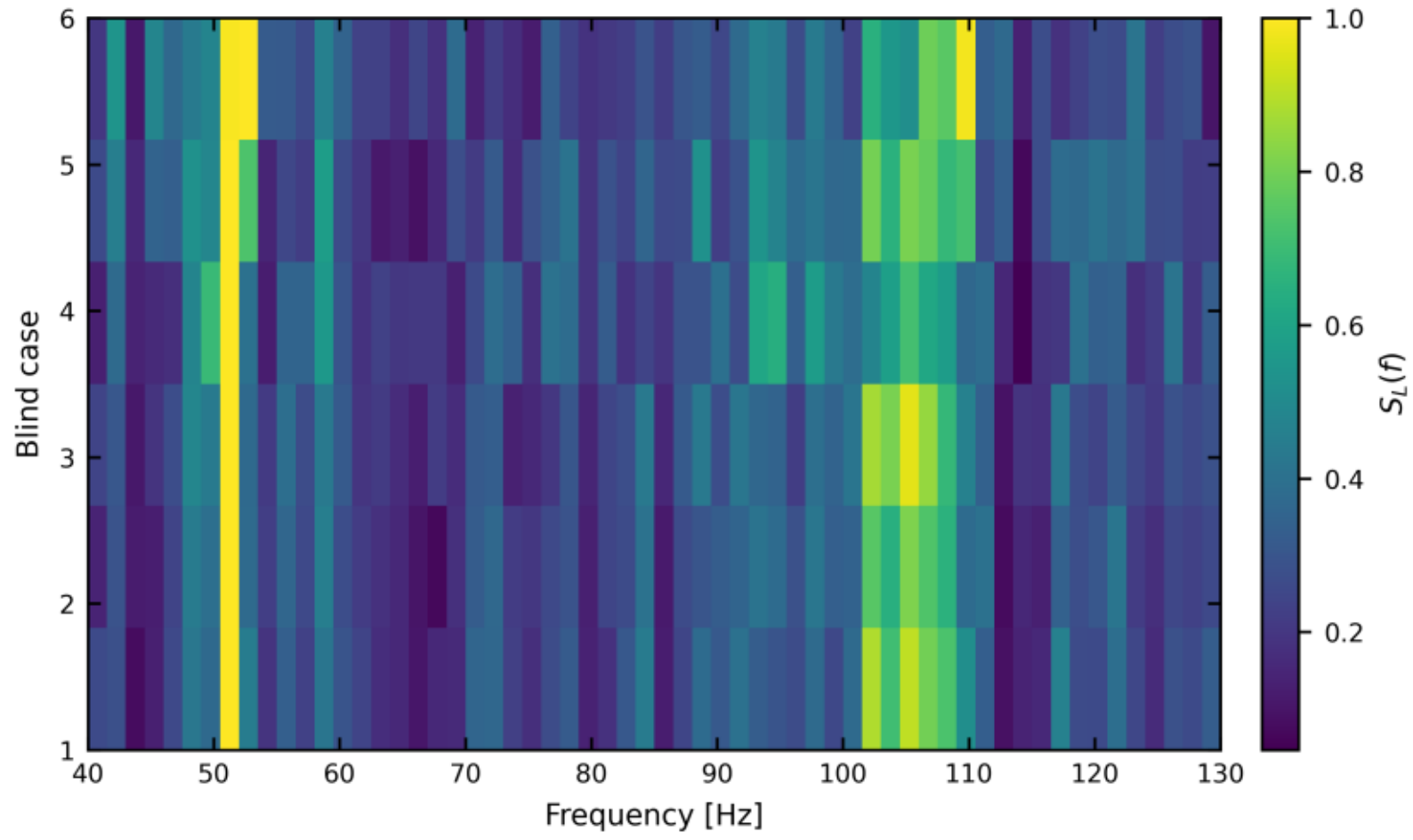

*Figure 8 Heatmap of normalized sensitivity to L across the blind cases*

The quantitative sensitivity statistics support the visual interpretation. The mean sensitivity to R over the blind cases is 0.305312 with a standard deviation of 0.184042, while the mean sensitivity to L is 0.341507 with a standard deviation of 0.179760. The slightly higher mean value for L indicates that, on average, the acceleration FRFs used in this experiment carry somewhat stronger information about inductance than resistance. This result is physically plausible because L is directly associated with the resonant behavior of the electrical branch, whereas R mainly controls the dissipative and damping contribution of the shunt circuit. In other words, changes in L tend to produce clearer frequency-localized shifts in the FRF, while changes in R modify attenuation and damping in a more distributed manner.

*Table 3 Summary statistics of the frequency-wise sensitivity analysis for R and L*

| Metric | R sensitivity | L sensitivity |
|---|---|---|
| Mean normalized sensitivity | 0.305312 | 0.341507 |
| Standard deviation of normalized sensitivity | 0.184042 | 0.179760 |
| Mean across-case standard deviation | 0.066548 | 0.074551 |
| Maximum across-case standard deviation | 0.236613 | 0.257595 |

The across-case standard deviation further shows that the sensitivity structure is stable across the blind configurations. For R, the mean across-case standard deviation is 0.066548 and the maximum is 0.236613. For L, the corresponding values are 0.074551 and 0.257595. The small mean values indicate that the model repeatedly uses similar frequency bands across the different blind cases. At the same time, the larger maximum values show that some localized frequencies become more or less important depending on the particular shunt pair. The slightly larger variability for L is also consistent with the stronger dependence of the electrical resonance position on the inductance value.

## Reduced Symbolic Distillation

The last stage of explainability involves applying symbolic regression to get an approximate equation for the reduced frequency-response descriptor. It is important to note that the purpose is not to reconstruct the governing equations for the coupled system but rather to describe the variation in the spectrum from the nominal SATMD system based on varying physical parameters.

The forward frequency-response evaluation involves evaluating the FRF of the system by passing the parameter vector [R, L, $m_2$, $k_p$] into the frequency-response solver. This comparison is compressed into one scalar descriptor, as defined in eq. (28). The dataset was built by sampling the four physical parameters R, L, $m_2$, and $k_p$ over the selected ranges. For every sampled combination, the FRF was computed and $\xi_{FRF}$ was extracted. This produced 6400 input-output samples of the form, $[R_i, L_i, m_{2,i}, k_{p,i}] \rightarrow \xi_{FRF,i}$

Symbolic regression was conducted using the PySR software package, which finds concise analytical models using the set of mathematical operators. Prior to PySR, normalization of the four input variables was performed to eliminate any numerical imbalance in the data. For this reason, the discovered expression is written using $x_0$, $x_1$, $x_2$, and $x_3$, corresponding respectively to normalized R, L, $m_2$, and $k_p$:

$$x_0 = \tilde{R}\,,\ x_1 = \tilde{L},\ x_2 = \widetilde{m_2},\ x_3 = \widetilde{k_p}\,, \qquad (29)$$

PySR was then used to search for an explicit expression approximating $\xi_{FRF} \approx f_{SR}\left(\tilde{R},\, \tilde{L},\, \widetilde{m_2},\, \widetilde{k_p}\right)$. The search used elementary binary operators (+, −, ×, /) together with simple unary operators. A complexity penalty was used so that PySR did not only minimize the prediction error but also favored expressions that remained compact and interpretable. The selected symbolic expression was $\xi_{FRF} \approx 1.69 + 0.02(x_3 - x_1 - 2.77 - 0.28x_2)[(x_1 - 3.14)^2 + x_0]$.

The identified expression comprises all four measured variables via their normalized values. The first factor $x_3 - x_1 - 2.77 - 0.28x_2$ considers stiffness-related variable $x_3$, inductance-related variable $x_1$, and mass-related variable $x_2$.

The second factor takes into account nonlinear relation to inductance and also includes the resistance-related variable $x_0$. Thus, there is no separation of the electric and mechanical variables in the derived symbolic expression since the latter describes the FRF characteristic using coupling effect between electric part R, L, and mechanical absorber $m_2$, $k_p$. This interpretation is logical because the behavior of the SATMD depends on the tuning of shunt circuit and absorber together.

The symbolic model resulted in training MAE equal to $8.49 \cdot 10^{-2}$ and test MAE of $8.82 \cdot 10^{-2}$. Respectively, the $R^2$ for training data was $0.874$ and $0.863$ for testing data. A test value of $R^2$ equal to roughly $0.86$ shows that the symbolic representation explains around 86% of the variance of the descriptor FRF, which means that the formula contains the majority of the main effects caused by parameters, but it fails to reproduce the details of the descriptor. Such behavior was expected because the complex FRF on the frequency range is summarized into a single scalar quantity and is then approximated by a compact symbolic formula.

The predicted vs true plot in Figure 9 supports this conclusion. The points follow the diagonal line well, which means that the symbolic formula gives a meaningful reduced surrogate for the FRF descriptor. There is some scatter, particularly for large descriptor values, which means that the compact formula is not a precise governing equation. Rather, it can be seen as an engineering symbolic approximation of the global deviation from the nominal FRF as a function of R, L, $m_2$, and $k_p$.

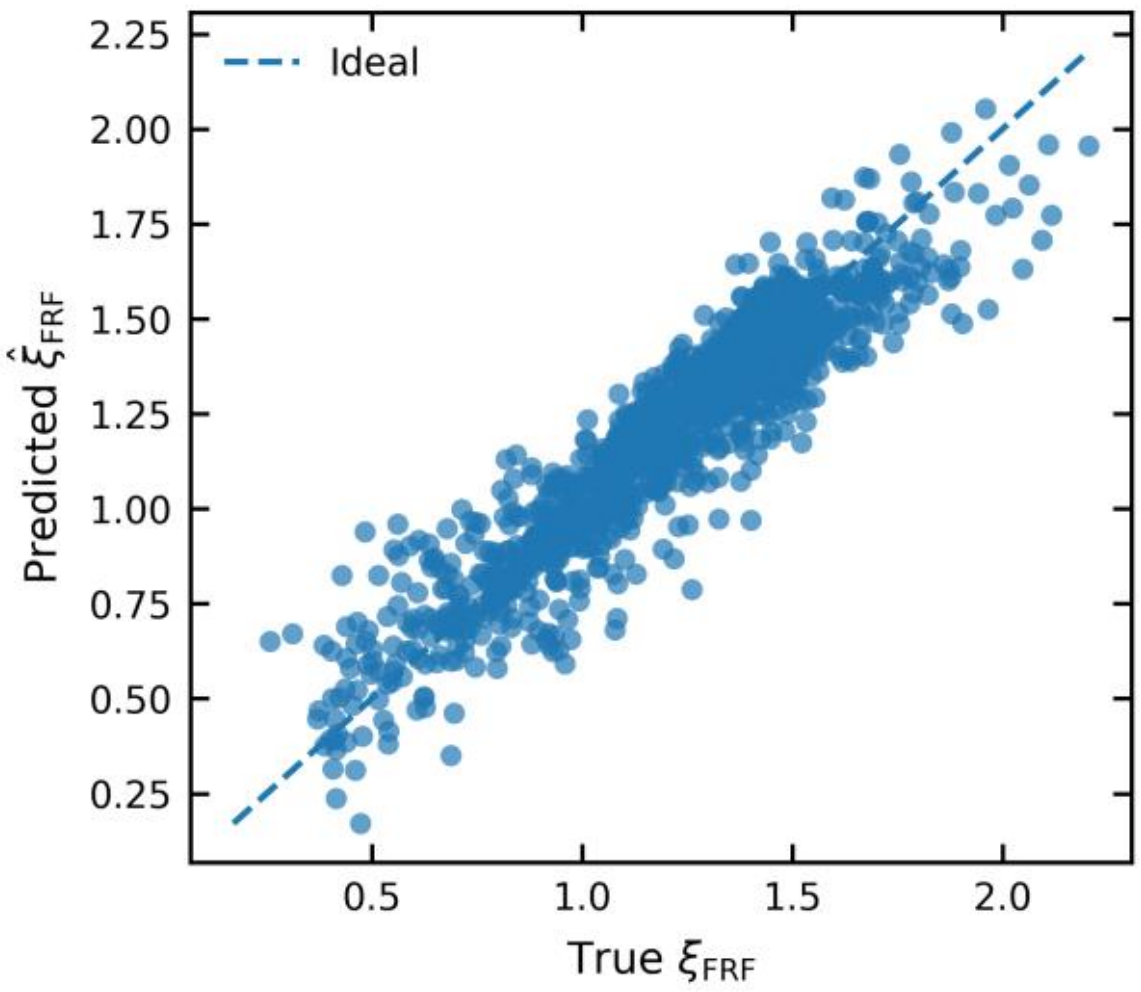

*Figure 9 Predicted versus true values of the global FRF descriptor using the PySR symbolic expression*

In summary, the reduced symbolic distillation process provides a compact surrogate for the dominant variation of the reduced FRF descriptor of the FRF data set produced. The forward solver is the reliable physical model, whereas the PySR equation gives us a concise mathematical model that explains the principal dependency of the global frequency response deviation.

## Discussion And Future Work

The results demonstrate three complementary capabilities of PI-NFRF. First, the differentiable frequency-domain solver enables physically interpretable parameter updating by minimizing the mismatch between complex target and reconstructed FRFs. The recovery of $m_2$, $k_p$, $R$, and $L$, together with the low FRF reconstruction error, verifies that the governing electromechanical model can be embedded successfully in a gradient-based identification loop. Nevertheless, a small reconstruction error does not by itself establish uniqueness of the identified parameters, since correlated parameter combinations may generate similar spectral responses. The parameter-identification results should therefore be interpreted as verification of the differentiable calibration procedure within the considered parameter range, rather than as general proof of global identifiability.

The second use case constitutes the main ML contribution of the framework. Instead of solving a separate optimization problem for every target FRF, the inverse network learns an amortized mapping from complex acceleration FRFs to the shunt parameters $R$and $L$. The low prediction errors obtained for the blind configurations indicate that the spectral response contains sufficient information for accurate inverse tuning within the sampled parameter region. Importantly, the network is not trained solely through parameter-label regression. The predicted shunt parameters are reintroduced into the frequency-domain solver and evaluated according to their ability to reconstruct the target FRFs. This closed-loop formulation preserves consistency between the learned inverse map and the underlying electromechanical model.

The reported errors remain small across the investigated shunt configurations, although they increase moderately toward the upper part of the sampled $R$-$L$ range. This behavior may be associated with lower sampling density near the boundaries of the training domain, changes in the conditioning of the inverse relationship, or reduced spectral distinction between neighboring parameter combinations. The current results therefore support interpolation within the investigated parameter space, while broader claims regarding extrapolation require additional evaluation outside the training range. A direct comparison with repeated physics-based optimization and a conventional supervised neural regressor would further quantify the benefit of the physics-consistent training strategy in terms of accuracy, robustness, and inference time.

The explainability analyses provide complementary insight into the behavior learned by the inverse model. The PCA projections show that the 64-dimensional penultimate feature representations are organized smoothly according to the sampled electromechanical configurations. However, because $R$ and $L$vary jointly in the considered blind cases, the observed curved organization should be interpreted as a combined electromechanical tuning coordinate rather than as complete disentanglement of the two shunt parameters. Independent sweeps in which $R$varies at fixed $L$, and vice versa, would be required to establish whether the network encodes their effects along distinct feature-space directions.

The frequency-wise gradient analysis indicates that the inverse predictions depend primarily on limited spectral regions near approximately 50–52 Hz and 103–110 Hz. These regions are associated with strong changes in resonance, anti-resonance, and electromechanical modal interaction. The gradients nevertheless quantify the local dependence of the neural outputs on the input FRFs and should therefore be interpreted as neural attribution rather than direct physical sensitivity. A stronger physical validation would compare these attribution maps with the derivatives of the forward-model response with respect to $R$and $L$. Agreement between the two would demonstrate that the network relies on frequency regions that are also physically informative for the corresponding shunt parameters.

The reduced symbolic-distillation analysis provides a compact approximation of the global normalized deviation from the nominal FRF. The resulting expression captures the combined influence of the mechanical and electrical parameters and confirms that the spectral response cannot generally be separated into independent mechanical and electrical contributions. However, the symbolic expression approximates a scalar reduction of the complete complex FRF and should not be interpreted as a recovered governing equation or as a replacement for the forward electromechanical solver. Its value lies in providing an interpretable engineering-level summary of the dominant parameter dependencies. Additional symbolic models based on resonance shifts, peak attenuation, anti-resonance location, or modal splitting could offer more direct physical interpretation.

The present study also has several limitations. The inverse model is trained using FRFs generated by a low-order electromechanical model, while validation is currently restricted to a limited number of blind configurations. Consequently, the effects of model-form discrepancy, sensor position, measurement noise, circuit-component tolerances, parasitic resistance, and unmodeled nonlinear behavior have not yet been assessed systematically. The joint variation of $R$and $L$also limits conclusions regarding their independent identifiability. In addition, the current results do not include comparisons with purely supervised neural networks, repeated optimization, or alternative spectral representations. Future work should therefore examine independently sampled shunt parameters, larger experimental datasets, uncertainty-aware inverse prediction, baseline and ablation studies, and transfer to more complex structural and electromechanical configurations.

## Conclusions

This work introduced PI-NFRF, an explainable physics-informed neural frequency-response framework for identification and inverse tuning of semi-active shunted piezoelectric tuned mass dampers. The framework combines a differentiable electromechanical frequency-domain solver, a neural inverse model, and explainability methods within a unified workflow.

The physics-informed identification stage accurately reconstructed the experimental complex FRF and recovered physically interpretable mechanical and electrical parameters. The inverse neural model, trained using synthetic FRFs and evaluated on six independently acquired experimental configurations, achieved mean absolute errors of $0.0823\ \Omega$for resistance and $5.72 \times 10^{-4}$ Hfor inductance, corresponding to mean relative errors of $0.41\%$and $0.66\%$, respectively. These results demonstrate successful synthetic-to-experimental transfer and enable rapid inference without solving a separate optimization problem for each measured response.

The explainability analyses showed that the learned feature representations vary systematically with the electromechanical tuning conditions. The neural attribution maps identified the frequency regions near $50$–$52$ Hz and $103$–$110$ Hz as particularly influential, while symbolic distillation provided a compact approximation of the global FRF deviation in terms of the mechanical and electrical parameters.

Overall, PI-NFRF provides a physically consistent and data-efficient approach for frequency-response-based identification and inverse tuning of semi-active electromechanical absorbers. The demonstrated transfer from synthetic training data to experimental FRFs indicates that the framework can reduce dependence on large labeled experimental datasets and avoid repeated optimization for each new measurement. Future work will focus on expanding the experimental dataset, independently varying $R$and $L$to assess their separate identifiability, and evaluating the method under different levels of measurement noise, sensor placement, circuit tolerances, and model-form uncertainty. Further developments will include comparisons with conventional supervised neural networks and repeated physics-based optimization, uncertainty-aware predictions with confidence intervals, and online adaptation to changes in structural or operating conditions. The framework will also be extended to higher-order and nonlinear smart structures, multimodal electromechanical absorbers, and real-time closed-loop retuning applications.

## Data and code availability

The underlying code and data for this study is available in github and be accessd via this link https://github.com/insane-group/Physics-Informed-Neural-Frequency-Responce-Framework